\documentclass[prb,twocolumn,superscriptaddress,floatfix]{revtex4-2}
\usepackage{amsfonts}
\usepackage{graphicx}
\usepackage{float}
\usepackage{amsmath}
\usepackage{amssymb}
\usepackage{hyperref}
\usepackage{pgf}
\usepackage{physics}
\usepackage{booktabs}
\usepackage{multirow}
\usepackage{makecell}
\usepackage{diagbox}
\usepackage[left=0.6in,right=0.6in,top=0.6in,bottom=0.6in
]{geometry}

\begin{document}

\title{Equivalence of residual entropy of hexagonal and cubic ices from
tensor network methods}
\author{Xia-Ze Xu}
\thanks{These authors contributed equally.}
\affiliation{Department of Physics, Tsinghua University, Beijing 100084, China}

\author{Tong-Yu Lin}
\thanks{These authors contributed equally.}
\affiliation{Department of Physics, Tsinghua University, Beijing 100084, China}

\author{Guang-Ming Zhang}
\email{zhanggm@shanghaitech.edu.cn}
\affiliation{State Key Laboratory of Quantum Functional Materials and School
of Physical Science and Technology, ShanghaiTech University, Shanghai 201210, China}
\affiliation{Department of Physics, Tsinghua University,
Beijing 100084, China}

\date{\today }

\begin{abstract}
The long-standing question of whether the residual entropy of hexagonal ice (%
$S_h$) equals that of cubic ice ($S_c$) remains unresolved despite decades
of research on ice-type models. While analytical studies have established
the inequality $S_h \geq S_c$, numerical investigations suggest that the two
values are very close. In this work, we revisit this problem using
high-precision tensor network methods. In the Monte Carlo approaches used
most commonly the residual entropy cannot be directly obtained by sampling
the ground-state degeneracy space. However, the tensor network framework
enables an explicit encoding of the ``ice rule'' into local tensors, and
then the residual entropy is transformed into finding the largest eigenvalue
of a transfer operator in the form of a projected entangled-pair operator,
which allows high-accuracy numerical evaluation. Meanwhile, we propose a
perspective based on analyzing the \textit{normality} of the transfer
operator and examine it numerically with variational tensor network methods.
This analysis allows for a direct computation for both residual entropies
with our recently developed split corner transfer matrix renormalization
group algorithm, providing a numerical evidence supporting the equality
between $S_h$ and $S_c$.
\end{abstract}

\maketitle

\section{Introduction}

The study of water ice has a long history and remains an active field of
research today. Although water ice is ubiquitous on Earth, it exhibits a
remarkably complex and rich phase diagram under varying temperature and
pressure conditions, along with several unusual properties compared with
ordinary solids~\cite{Bartels-Rausch_2012, Bartels_2013, Salzmann_2019}. One
of these anomalous properties is the \textit{residual entropy}, where water
ice retains a finite entropy even at temperatures well below the freezing
point.

\begin{figure}[t]
\centering
\includegraphics[width=1\linewidth]{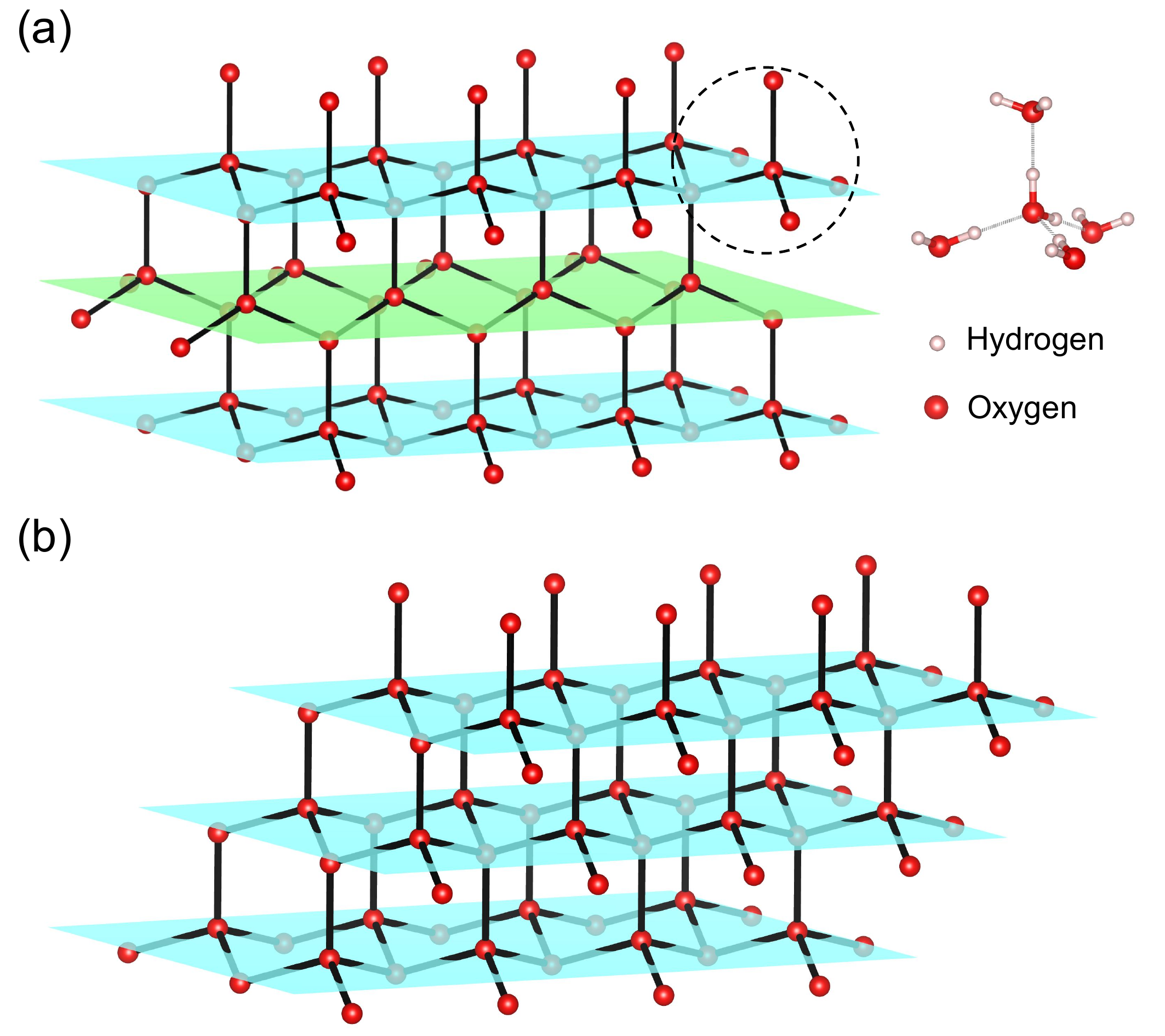}
\caption{Structure of hexgonal ice $I_h$ (a) and cubic ice $I_c$ (b). The
cyan (green) layer denotes the crinkled honeycomb layer (and its mirror
counterpart). The dashed circle marks a tetrahedral unit, and one of the
hydrogen configurations satisfying the ice rule is shown to the right.}
\label{Fig:Structure of ice}
\end{figure}

The microscopic origin of this residual entropy was first elucidated by
Pauling~\cite{pauling1935}. At that time, the oxygen sublattice of ice was
clear \cite{Bragg_1921}. As shown in Fig.~\ref{Fig:Structure of ice}(a), the
oxygen atoms form a wurtzite lattice in which each oxygen atom is
tetrahedrally coordinated with four nearest-neighbor oxygen atoms. Based on
this structural model, known as hexagonal ice ($I_h$), Pauling proposed that
a hydrogen atom is located between each pair of neighboring oxygen atoms,
positioned closer to one oxygen atom and farther from the other. Then he
assumed that the hydrogen atoms in $I_h$ are disordered and that all
configurations satisfy the so-called \textit{Bernal-Fowler ice rules}\cite{bernal_1933},
which for each oxygen atom two of the four neighboring hydrogens are ``close"
and two are ``far" are equally probable. Assuming that the allowed hydrogen configurations around each oxygen are
independent of one another---an approximation known as \textit{Pauling's
argument}---he estimated the number of configurations satisfying the ice
rule. He found that this number is extensive in the number of water
molecules and can be approximated as
\begin{equation}
W_h \approx \left( \frac{3}{2} \right)^N,
\end{equation}
where $W_h$ is the total number of configurations and $N$ is the number of
molecules. This leads to a residual entropy per molecule of
\begin{equation}
S_h = \lim_{N \to \infty} \frac{1}{N} \ln W_h = \ln(w_h)\approx \ln(\frac{3}{2}),
\end{equation}
where $w_h$ is the number of configurations per site. This result is
consistent with experimental observations \cite{Giauque_1933,Gordon_1934}.

\begin{table*}[t]
\caption{Residual entropy of hexagonal and cubic ice obtained using
different methods. The results are expressed in the form of configuration
number per site $w=e^{S}$.}%
\begin{ruledtabular}
\begin{tabular}{llll}\label{Table: MC results}
Methods&Group and year&$w_h$ (hexagonal ice)&$w_c$ (cubic ice)\\
\hline
Series expansion &Nagle (1966) \cite{nagle1966} & 1.50685(15) &1.50685(15)\\
Multicanonical algorithm &Berg \textit{et al.} (2012) \cite{berg2012}  &1.507117(35) & \\
Thermodynamic integration &Herrero  and Ramirez (2014) \cite{Herrero_2014} &1.50786(12) &1.50778(11)\\
&Kolafa (2014) \cite{kolafa2014}& 1.5074674(38) & 1.5074660(36)\\
Wang–Landau algorithm &Ferreyra  and Grigera (2018) \cite{ferreyra2018}&1.5070(9) & 1.50694(12)\\
&Hayashi \textit{et al.} (2021) \cite{Hayashi_2021}&1.5074723(474)&\\
&Li \textit{et al.} (2022) \cite{Li2022} & 1.507427(23) & 
\end{tabular}
\end{ruledtabular}
\end{table*}

Ever since the pioneering work of Pauling, the study of the residual entropy
in ice-type systems has evolved into a canonical problem in statistical
physics. In general, on any regular lattice with a coordination number of
4, one can assign a binary degree of freedom analogous to the ``near''
and ``far'' hydrogen positions in water ice, and count the number of
configurations in which the degrees of freedom on the edges around each
vertex satisfy the ice rule. This problem has been extended to other
lattices, including two-dimensional (2D) systems~\cite%
{lieb1967,Lieb_1967,lin_1976,lin_1983,kirov_2012,KIROV_2013,Li_2023}, as
well as various three-dimensional (3D) lattices~\cite%
{nagle1966,kolafa2014,li2024,Herrero_2014}.

In efforts to estimate the residual entropy of ice-type models, a variety of
theoretical and numerical methods have been proposed and developed. On the
theoretical side, Onsager proved that Pauling's argument provides a lower
bound for the entropy \cite{Onsager_1960}. In addition, series-expansion
techniques have been introduced to improve the estimation \cite%
{DiMarzrio_1964, nagle1966}. A major advance was achieved by Lieb, who
obtained an exact result for the 2D square-ice model using the
transfer-matrix method\cite{lieb1967,Lieb_1967}. On the numerical side,
Monte Carlo simulations are among the most commonly used approaches,
including thermodynamic integration methods\cite%
{Isakov_2004,Herrero_2013,kolafa2014,Herrero_2014}, multicanonical
simulations~\cite{Berg_2007,berg2012}, and Wang-Landau algorithms\cite%
{ferreyra_2016,ferreyra2018,Li2022,Andriushchenko_2019,Hayashi_2021}.

Despite the long history of studies on ice-type models, one of the
long-standing unsolved problems concerns whether the residual entropy of
hexagonal ice ($S_h$) equals that of cubic ice ($S_c$). Cubic ice ($I_c$)
possesses a diamond-cubic lattice structure consisting of a repeating
face-centered cubic array of tetrahedrally coordinated water molecules, as
illustrated in Fig.~\ref{Fig:Structure of ice}(b). Onsager first noted the
relationship between $S_h$ and $S_c$. Analytically, he proved that there
exists an inequality between the residual entropies of $I_h$ and $I_c$~\cite%
{nagle1966,li2024}:
\begin{equation}
S_h \geq S_c.
\end{equation}

However, those results from series expansions and Monte Carlo simulations
suggest that these two values are extremely close to each other, as
summarized in Table~\ref{Table: MC results}. This problem is also
experimentally relevant, as $I_h$ is the most common ice structure in
nature, while pure solid $I_c$ has also been discovered recently~\cite%
{komatsu_2020,del_2020,huang_2023}.

In this paper, we aim to address this long-standing problem using the
powerful framework provided by tensor network methods, which have been shown
to yield high-precision estimates of residual entropy \cite%
{Vanderstraeten2018,Vanhecke2021,colbois_2021,colbois_2022a,song2023a}.
In the Monte Carlo methods, the residual entropy
cannot be directly obtained from sampling the ground-state degeneracy space,
but usually inferred through numerical integration of the heat capacity over
finite temperature, while the tensor network approach has the
advantage of directly representing and computing the residual entropy in the
ground-state degeneracy space. Within the tensor network framework, the ice
rule on each vertex can be naturally encoded in a local tensor, and the 
number of configurations can be represented as a tensor network defined on
the same geometry as the original lattice. The residual entropy can then be
computed analogously to the transfer operator formalism in statistical
physics, in which the partition function problem is transformed into solving
for the largest eigenvalue of a transfer operator. In our case, both the
transfer operator and its eigenvectors are represented in tensor network
form as a tensor network operator and a tensor network state,
respectively.

The residual entropies of $I_h$ and $I_c$ have been previously studied using
variational tensor network methods~\cite{Vanderstraeten2018}. However, that
approach can only be safely applied to Hermitian transfer operators, whereas
the transfer operator $\hat{M}$ of $I_c$ is non-Hermitian. To circumvent
this issue, the authors of the paper \cite{Vanderstraeten2018} imposed an
additional symmetry constraint on the leading eigenvectors to enforce
Hermiticity. Under this constraint, the entropies of $I_h$ and $I_c$ became
indistinguishable.

In contrast, in this work we address the problem from a different
perspective by investigating the \textit{normality} of the transfer operator
$\hat{M}$. As proposed in a recent study \cite{Tang2025}, normality of the
transfer operator is sufficient to ensure the tensor network method remains
applicable to non-Hermitian transfer operators. In our context, the
relevance of normality is twofold. (1) If $\hat{M}$ is strictly normal, one
can rigorously prove that $S_h = S_c$. (2) Even if not strict, a high degree
of normality, which could be measured numerically, can provide a theoretical
basis for applying variational tensor network methods without imposing
additional symmetry constraints, thereby enabling a direct comparison
between $S_h$ and $S_c$.

The rest of this paper is organized as follows. In Sec.~\ref{Sec:: Tensor
network representation for residual entropy problem}, we present the tensor
network representation of the residual-entropy problem. In Sec.~\ref{Sec::
Normality}, we explain the concept of normality for the transfer operator
and measure it numerically using tensor network techniques. We then
demonstrate the direct calculation of the residual entropies of $I_c$ and $%
I_h$ using our recently developed split corner transfer matrix
renormalization group method~\cite{Xu2025} in Sec.~\ref{Sec:: Variational
tensor network method for the residual entropy}. Finally, Sec.~\ref{Sec::
Conclusion and discussion} provides a summary and discusses possible future
extensions of this work.

\section{Tensor network representation of the residual entropy problem}

\label{Sec:: Tensor network representation for residual entropy problem}

\begin{figure}[t]
\centering
\includegraphics[width=1\linewidth]{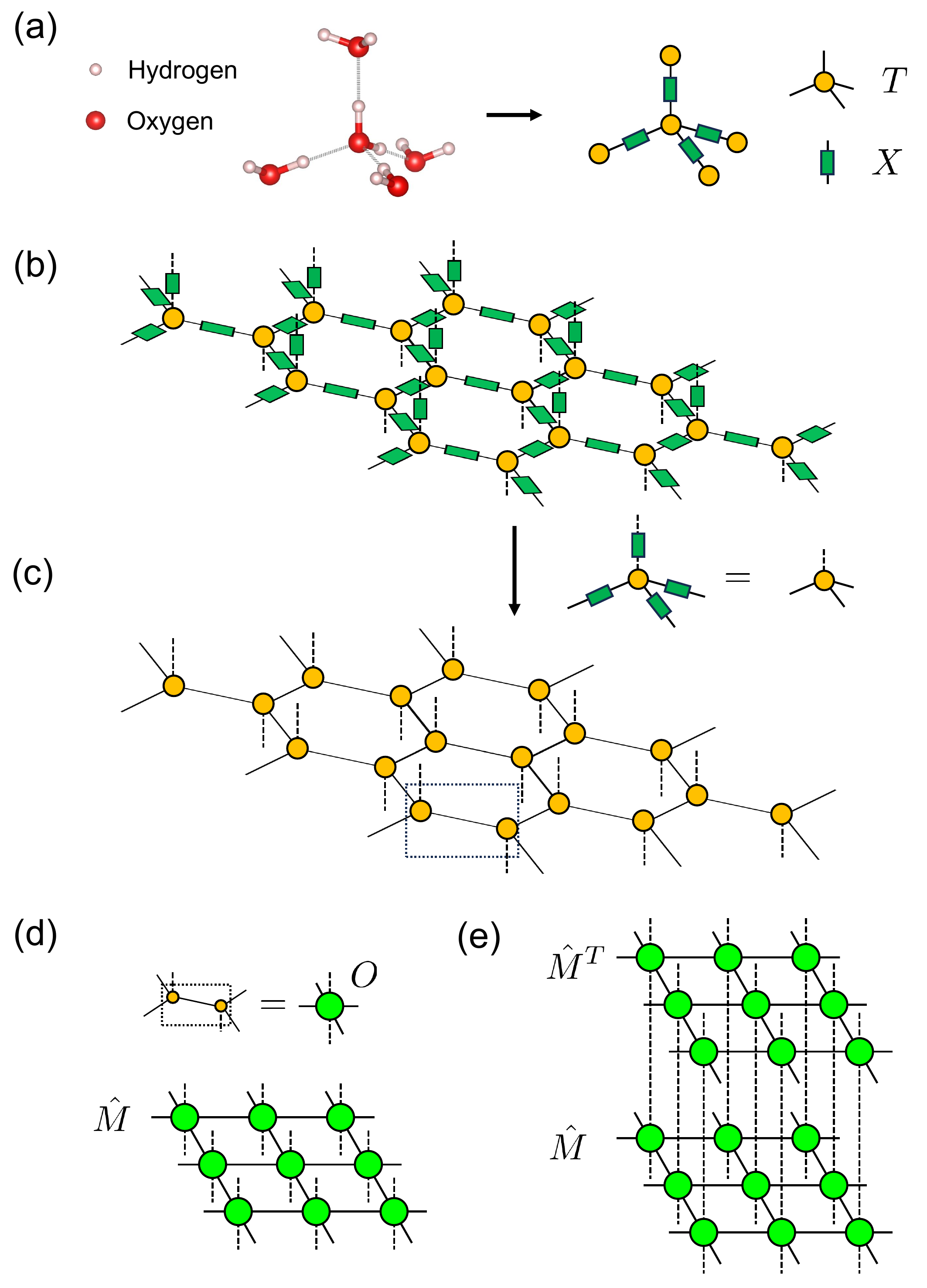}
\caption{Tensor network representation of the residual entropy problem for
ice $I_h$ and $I_c$. (a) Local tetrahedral structure of water molecules and
the corresponding local tensors encoding the ice rule. (b) Tensor network
representation for the layer-to-layer transfer operator. Dashed black lines
denote intralayer bonds. (c) Simplified representation obtained by absorbing
the $X$ matrix into $T$ tensors on one of the two sublattices of the
honeycomb lattice. (d) Transfer operator for $I_c$, consisting of a uniform
local tensor $O$, where $O$ is built by blocking neighboring $T$ tensors.
(e) Transfer operator for $I_h$.}
\label{Fig:transfer matrix construction}
\end{figure}

The number of configurations satisfying the ice rule can be represented
naturally in the tensor network language. The tensor network representation
of hexagonal ice and cubic ice has been introduced previously \cite%
{Vanderstraeten2018} and we review it here, focusing on the relation between
hexagonal and cubic ice.

First, we elaborate on the tensor network representation of the residual
entropy of cubic ice. We begin by illustrating how the ice rule can be
encoded into a local tensor. As shown in Fig.~\ref{Fig:transfer matrix
construction}(a), the local tetrahedral geometry of oxygen atoms can be
represented by a four-legged tensor $T$. Each bond has a dimension of 2,
corresponding to the ``near'' (index 0) and ``far'' (index 1) positions of
the hydrogen atoms. The constraints are implemented using two types of
tensors: (1) a Pauli $X$ matrix on each link ensures the presence of one
hydrogen atom per bond, and (2) the ``2 close, 2 far'' ice rule is encoded
in the local vertex tensor $T$, defined as
\begin{equation}
T(s_1,s_2,s_3,s_4) =
\begin{cases}
1, & \text{if exactly two of the } s_i = 0, \\
0, & \text{otherwise},%
\end{cases}%
\end{equation}
where the statistical weight of a hydrogen arrangement satisfying the ice
rule is set to 1, and the arrangements violating the rule are assigned a
weight of 0.

With the local tensor structure established, the residual entropy of cubic
ice can be mapped to a tensor network. Since the oxygen lattice structure
can be regarded as a stacking of crinkled honeycomb layers, we focus first
on the tensor network representation of a single layer. As illustrated in
Fig.~\ref{Fig:transfer matrix construction}(b), the crinkled honeycomb
lattice can be directly mapped onto a tensor network, where $T$ tensors
occupy the honeycomb sites, Pauli $X$ matrices reside on the intralayer
bonds, and open indices represent interlayer connections.

\begin{figure*}[t]
\centering
\includegraphics[width=1\linewidth]{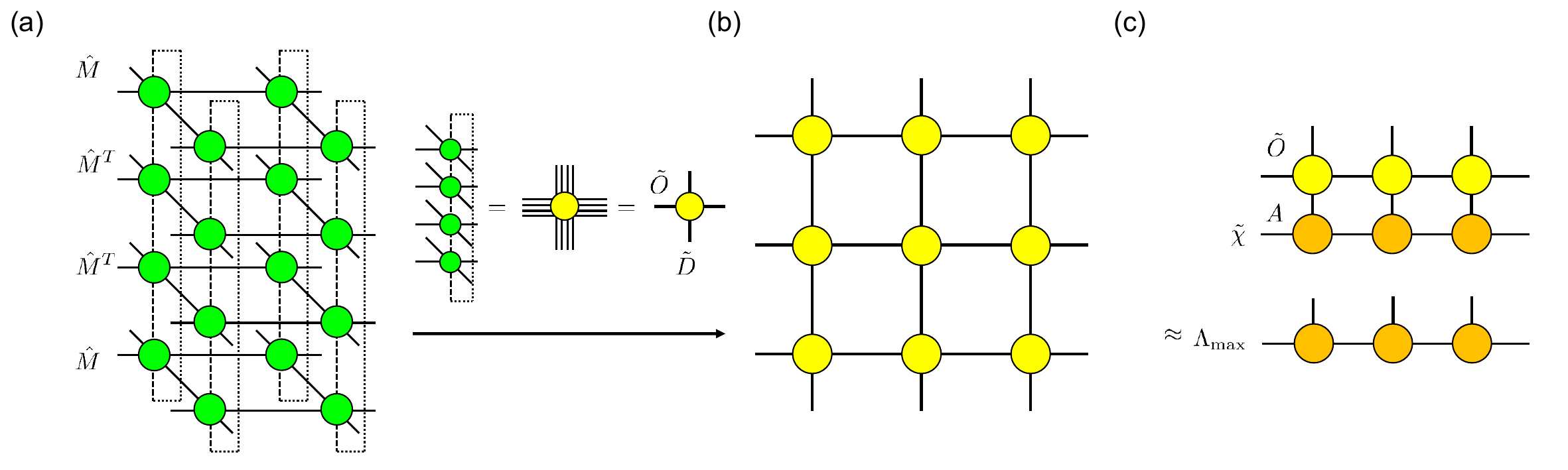}
\caption{Tensor network method for calculating the measure of normality. (a)
Tensor network representation for the numerator in Eq.~
\eqref{eqn:fidelity_definition}. (b) The network is transformed into an
equivalent 2D tensor network consisting of uniform tensor $\tilde{O}$ by
tracing the vertical bonds. (c) Contracting the 2D tensor network with
boundary MPS $\ket{\psi(A)}$.}
\label{Fig:Normality}
\end{figure*}

The network can be further simplified by observing that the tensor $T$ is
invariant under the action of four $X$ matrices:
\begin{equation}
\sum_{\{s_i\}} \left(\prod_{i=1}^4X(s_i^{\prime
},s_i)\right)T(s_1,s_2,s_3,s_4)= T(s_1^{\prime },s_2^{\prime },s_3^{\prime
},s_4^{\prime }).
\end{equation}
Since the lattice is bipartite, the Pauli $X$ matrices on links can be
absorbed into one of the two sublattices. As a result, the network reduces
to a form composed solely of $T$ tensors, as illustrated in Fig.~\ref%
{Fig:transfer matrix construction}(c). By further grouping pairs of adjacent
$T$ tensors into a single tensor $O$, the single-layer network can be
represented as a tensor network operator $\hat{M}(O)$, as shown in Fig.~\ref%
{Fig:transfer matrix construction}(d). The full 3D network is finally
obtained by repeatedly stacking the operator $\hat{M}(O)$ along the vertical
direction.

For hexagonal ice, the construction follows a similar procedure. The primary
difference is that the lattice is formed by stacking crinkled honeycomb
layers in an $ABAB$ pattern, where layer $B$ is the mirror image of layer $A$%
. Consequently, the 3D tensor network is constructed by repeatedly stacking
the double-layer structure, as shown in Fig.~\ref{Fig:transfer matrix
construction}(e), with $\hat{M}^T$ denoting the transpose of the operator $%
\hat{M}$.

The residual entropy problem of water ice can be reformulated in terms of a
3D tensor network, which can be written as:
\begin{equation}
W=\text{Tr}\{\hat{T}^{N_z}\},
\end{equation}
where $\hat{T}$ denotes the layer-to-layer transfer operator composed of a
layer of local tensors, and $N_z$ is the number of layers. For cubic ice we
have $\hat{T}_c=\hat M$, whereas for hexagonal ice the transfer operator
takes the form $\hat{T}_h=\hat M\hat M^T$.

Contracting the 3D tensor network will give the total number of
configurations satisfying the ice rule. In the thermodynamic limit, this
contraction reduces to computing the dominant eigenvalue of the transfer
operator,

\begin{equation}
\hat{T}\ket{\psi}=\lambda_{\text{max}} \ket{\psi},
\label{eqn: fixed-point equation}
\end{equation}
from which the configuration number per site is obtained as
\begin{equation}
\begin{aligned} w_c&=\lim_{N\to
\infty}\left(\lambda_{\hat{M},\text{max}}\right)^{\frac{1}{N}},\\
w_h&=\lim_{N\to
\infty}\left(\lambda_{\hat{M}\hat{M}^T,\text{max}}\right)^{\frac{1}{2N}},
\end{aligned}
\end{equation}
where $N$ is the number of sites within a single honeycomb layer.

\section{Normality analysis for the transfer operator}

\label{Sec:: Normality}

With the tensor network representation of the transfer operator for $I_c$
and $I_h$ established, we proceed to investigate the properties of the
transfer operator $\hat{M}$. Although it can be readily shown that $\hat{M}$
is non-Hermitian, an important but previously unexplored property is its
\textit{normality}. Recall from linear algebra that a matrix $\hat{M}$ is
said to be normal if
\begin{equation}
\hat{M} \hat{M}^T = \hat{M}^T \hat{M}.
\end{equation}
Moreover, if $\hat{M}$ is normal, the eigenvalues of $\hat{M}\hat{M}^T$ are
in one-to-one correspondence with the squared norms of the eigenvalues of $%
\hat{M}$. Since $\hat{M}$ is, by construction, also a non-negative matrix,
the Perron-Frobenius theorem guarantees the existence of a real leading
eigenvalue. Combining these two properties yields
\begin{equation}
\lambda_{\hat{M}\hat{M}^T,\mathrm{max}} = \lambda_{\hat{M},\mathrm{max}}^2,
\end{equation}
and consequently, $S_h = S_c$.

Although the analytical treatment of projected entangled pair operators
(PEPOs) is notoriously difficult, the normality of $\hat{M}$ can be checked
numerically by computing the measure of normality defined as~\cite{Tang2025}%
\begin{equation}
\mathcal{N}(\hat{M}) = \mathcal{F}(\hat{M}\hat{M}^T, \hat{M}^T \hat{M}),
\end{equation}
where $\mathcal{F}$ quantifies the fidelity per site between two operators
and is given by
\begin{equation}  \label{eqn:fidelity_definition}
\mathcal{F}(A,B) = \lim_{N_{xy} \to \infty} \left( \frac{\Tr(A^\dagger B)}{%
\sqrt{\Tr(A^\dagger A)} \sqrt{\Tr(B^\dagger B)}} \right)^{\frac{1}{N_{xy}}},
\end{equation}
with $N_{xy}$ denoting the number of local tensors comprising the operator
in the $xy$ plane.

All terms in Eq.~\eqref{eqn:fidelity_definition} can be represented as
tensor networks. For instance, the numerator can be expressed as a
four-layer network, as shown in Fig.~\ref{Fig:Normality}(a). By tracing out
the vertical legs, this network is reduced to an equivalent 2D tensor
network composed of local tensors $\tilde{O}$ with bond dimension $\tilde{D}
= 16$, as illustrated in Fig.~\ref{Fig:Normality}(b). To contract this 2D
network, we employ the variational uniform matrix product state (VUMPS)
method~\cite{haegeman2017,zauner-stauber2018,vanderstraeten2019}, which
allows us to solve for the leading eigenvector of the transfer operator $%
\hat{T}(\tilde{O})$:
\begin{equation}
\hat{T}(\tilde{O}) \ket{\psi(A)} = \Lambda_{\mathrm{max}} \ket{\psi(A)},
\end{equation}
using a uniform MPS $\ket{\psi(A)}$ with bond dimension $\tilde{\chi}$ as
the variational ansatz, as depicted in Fig.~\ref{Fig:Normality}(c).

\begin{table}[b]
\caption{Measure of normality of the transfer operator with different MPS
bond dimension $\tilde{\protect\chi}$. The fidelity between $\hat{M}$ and $%
\hat{M}^T$ is shown for reference.}%
\begin{ruledtabular}
\begin{tabular}{ccc}\label{Table: Normality result}

$\tilde{\chi}$ & $\mathcal{F}(\hat{M}\hat{M}^T,\hat{M}^T\hat{M})$ & $\mathcal{F}(\hat{M},\hat{M}^T)$ \\
\hline
20  & 0.9999137 & 0.9743303 \\
40  & 0.9999076 & 0.9743263 \\
60  & 0.9999060 & 0.9743258 \\
80  & 0.9999046 & 0.9743255 \\
100 & 0.9999042 & 0.9743254 \\
150 & 0.9999039 & 0.9743253 \\
200 & 0.9999039 & 0.9743253 \\
250 & 0.9999038 & 0.9743252 \\

\end{tabular}
\end{ruledtabular}
\end{table}

\begin{figure}[t]
\centering
\includegraphics[width=1\linewidth]{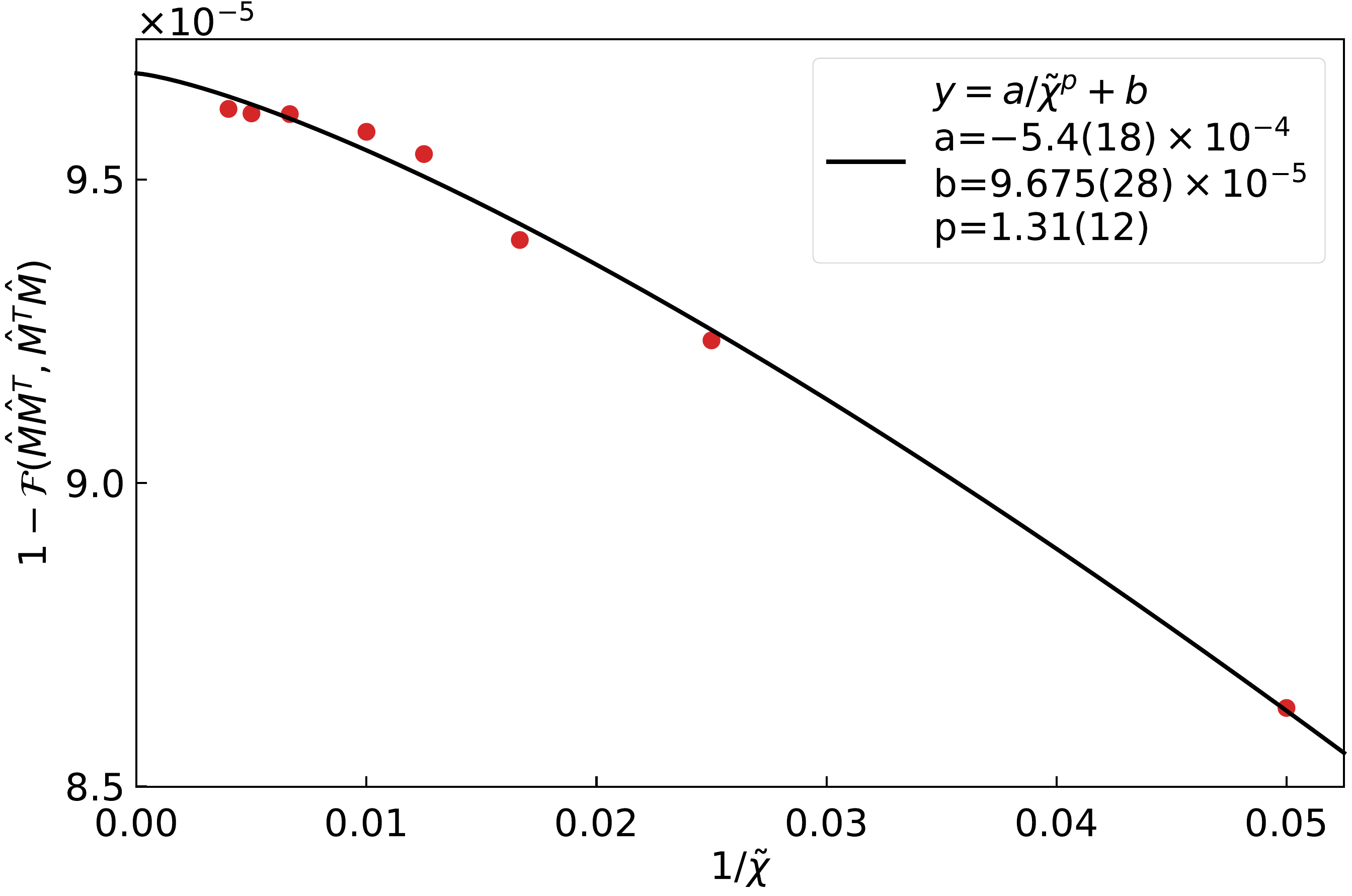}
\caption{Extrapolation of the normality measure $\mathcal{F}(\hat{M}\hat{M}%
^T,\hat{M}^T\hat{M})$ using the fitting form $1-\mathcal{F}(\tilde{\protect%
\chi})=a/\tilde{\protect\chi}^p+b$.}
\label{Fig:Normality_result}
\end{figure}

The results are shown in Table~\ref{Table: Normality result}, compared to
the fidelity between $\hat{M}$ and $\hat{M}^T$. We have calculated the
measure of normality $\mathcal{N}(\hat{M})$ to bond dimensions up to $\tilde{%
\chi}=250$ and the results show little change for $\tilde{\chi}\geq100$. As
shown in Fig.~\ref{Fig:Normality_result}, an extrapolation to $\tilde{\chi}%
\to\infty$ with fitting function $\mathcal{F}(\tilde{\chi})=a/\tilde{\chi}^p+%
\mathcal{F}_\infty$ gives $\mathcal{F}_\infty=0.99990325(28)$, demonstrating
that the transfer operator $\hat{M}$ is extremely close to a normal matrix,
although with a tiny violation.

Notice that normality of $\hat{M}$ is a sufficient but not necessary
condition for the equivalence of $S_h$ and $S_c$. Normality ensures all
eigenvalues of $\hat{M}\hat{M}^T$ and $\hat{M}$ to be in one-to-one
correspondence, while $S_h=S_c$ only requires the relation Eq.(10) between the largest eigenvalues.
In this sense, the high degree of normality of $\hat{M}$ provides a 
perspective to understand the relation between these two residual entropies,
but the small violation does not disprove the equality of $S_h$ and $S_c$.
To further elaborate the equivalence between $S_h$ and $S_c$, we will
perform a direct numerical calculation and compare these two residual
entropies in the next section, however, the analysis of normality in this section
is crucial for the application of variational optimization on solving the
leading eigenvector of non-Hermitian transfer matrix $\hat{M}$.

\section{Variational tensor network method for the residual entropy}

\label{Sec:: Variational tensor network method for the residual entropy}

The high degree of normality of the transfer operator $\hat{M}$ actually makes
it possible to calculate the residual entropy directly under thermodynamic
limit with the variational tensor network method. In this section, we
briefly introduce the variational tensor network approach, explain the
significance of normality, and present the residual entropy results obtained
using the  split corner transfer matrix
renormalization group (split-CTMRG) scheme.

In the tensor network method, the dominant eigenvector of the transfer
operator $\hat{T}$ contracted by local tensor $O$ is approximated by an
infinite projected entangled pair state (iPEPS) $\ket{\psi(A)}$, which
consists of a uniform local tensor $A$, as shown in Fig.~\ref{Fig:iPEPS
fixed point}. For the transfer operator of hexagonal ice $\hat {T_h}=\hat{M}%
\hat{M}^T$, which is Hermitian, the leading eigen equation (Eq.~\eqref{eqn:
fixed-point equation}) can be recast as a variational optimization problem
that maximizes
\begin{equation}
S(A,\overline{A})=\frac{1}{N_{xy}}\ln (\lambda )=\frac{1}{N_{xy}}\ln (\frac{%
\bra{\psi(\overline{A})}\hat{T}\ket{\psi(A)}}{\bra{\psi(\overline{A})}%
\ket{\psi(A)}}),  \label{eqn: residual entropy iPEPS expression}
\end{equation}
where $N_{xy}$ is the number of uniform tensors in each plane. $\frac{%
\bra{\psi}\hat{T}\ket{\psi}}{\bra{\psi}\ket{\psi}}$ is known as the Rayleigh
quotient in the mathematical context, and the cost function has the physical
meaning of the residual entropy.

\begin{figure}[t]
\centering
\includegraphics[width=1\linewidth]{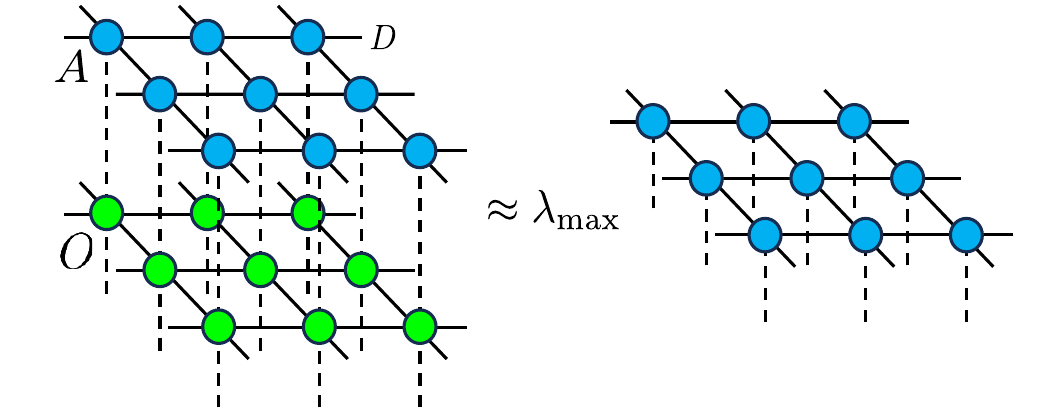}
\caption{The fixed point equation for transfer operator $\hat{T}$ with iPEPS
$\ket{\psi(A)}$ as an ansatz.}
\label{Fig:iPEPS fixed point}
\end{figure}

In contrast, the transfer operator of cubic ice $\hat{T_c}=\hat{M}$, is
non-Hermitian. Consequently, a variational principle of the form in Eq.~%
\eqref{eqn: residual entropy iPEPS expression} does not generally apply, and
the dominant eigenvalue can not be obtained from this expression. To keep
the variational principle valid for cubic ice, the previous work \cite%
{Vanderstraeten2018} restricted the variational manifold to iPEP, which is
invariant under a spatial $180^{\circ}$ rotation. Within this symmetric
subspace, $\hat{M}$ and $\hat{M}^T$ act identically, effectively restoring
the Hermiticity of $\hat{M}$. However, this symmetry constraint
simultaneously reduces the transfer operator of the hexagonal ice to the
form $\hat{M}^2$. As a result, within the constrained subspace, the residual
entropies of hexagonal and cubic ice become indistinguishable.

Instead, we show that the variational principle for the transfer operator $%
\hat{M}$ can be applied without imposing symmetry constraints on the iPEPS
manifold. If the transfer operator $\hat{T}$ is normal, eigenvalue
decomposition exists, so that any vector $\ket{\psi}$ can be expanded in the
basis of eigenvectors as
\begin{equation}
\ket{\psi}=\sum_i a_i\ket{\phi_i},\quad \sum_i |a_i|^2=1, \quad \hat{T}%
\ket{\phi_i}=\lambda_i\ket{\phi_i}.
\end{equation}
The Rayleigh quotient is then bounded by the dominant eigenvalue:
\begin{equation}
\left|\frac{\bra{\psi}\hat{T}\ket{\psi}}{\bra{\psi}\ket{\psi}}%
\right|=\left|\sum_i |a_i|^2\lambda_i\right|\leq\sum_i
|a_i|^2\left|\lambda_i\right|\leq\lambda_{\text{max}}.
\end{equation}
The upper bound is achieved when $\ket{\psi}$ is chosen to be the dominant
eigenvector. If the transfer operator $\hat{T}$ is also non-negative, the
Perron-Frobenius theorem ensures that its dominant eigenvalue is real and
that the eigenvector can also be taken to be real. Consequently, the
dominant eigenvector can be optimized within a real vector space and the
cost function remains real during the optimization.

\begin{figure}[t]
\centering
\includegraphics[width=1\linewidth]{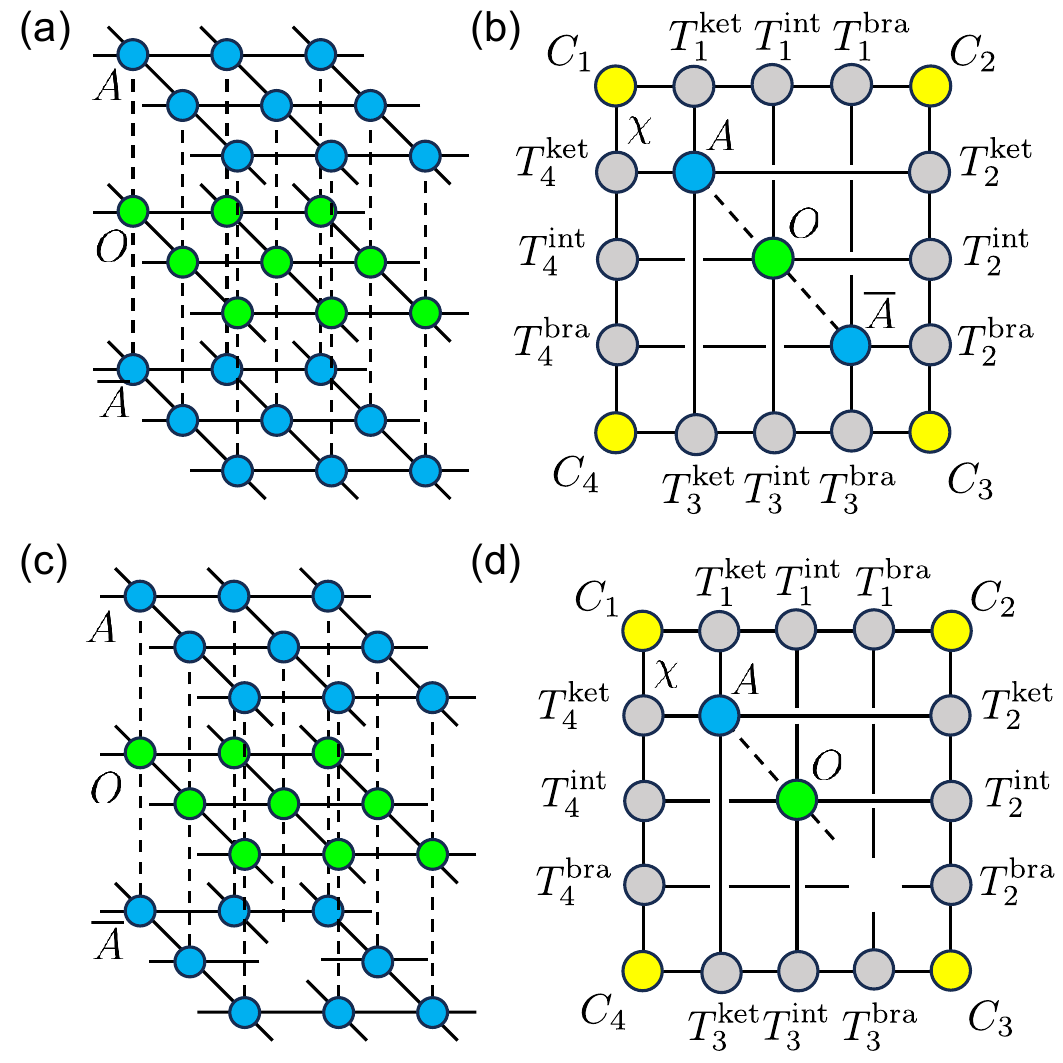}
\caption{Variational tensor network scheme and split-CTMRG method. (a),(c)
The triple-layer network structure for the terms involved in calculating
residual entropy and the corresponding gradient. (b),(d) The contraction of
the network in (a),(c) with split-CTM environment tensors.}
\label{Fig:Variational scheme}
\end{figure}

By combining these observations, we conclude that optimizing the Rayleigh
quotient of a non-negative and normal operator over the space of real
vectors yields the exact dominant eigenvalue. For operators that are not
strictly normal, numerical results reported in \cite{Tang2025} for 1D matrix
product operators (MPO) indicate that a sufficiently high degree of
normality, even if not strict, can ensure the effectiveness of variational
tensor network methods. Extending to the 2D PEPO case, this suggests that in
practice the direct variational optimization remains applicable for the
transfer operator of cubic ice $\hat{M}$, which is non-negative and highly
normal as analyzed in Sec.~\ref{Sec:: Normality}.

As the variational principles for both transfer matrices are established, we
proceed to present how to optimize the cost function using variational
tensor network method. The gradient of cost function $S(A,\overline{A})$
with respect to $A$ is given by
\begin{eqnarray}
g&=&2\times \partial _{\overline{A}}S(A,\overline{A})  \notag
\label{eqn: gradient expression} \\
&=&\frac{2}{N_{xy}}\left( \frac{\partial _{\overline{A}}\bra{\psi(%
\overline{A})}\hat{T}\ket{\psi(A)}}{\bra{\psi(\overline{A})}\hat{T}%
\ket{\psi(A)}}-\frac{\partial _{\overline{A}}\bra{\psi(\overline{A})}%
\ket{\psi(A)}}{\bra{\psi(\overline{A})}\ket{\psi(A)}}\right).  \notag \\
\end{eqnarray}
Both the cost function and gradient can be represented as an infinite 2D
tensor network. As shown in Figs.~\ref{Fig:Variational scheme}(a) and~\ref{Fig:Variational scheme}(b),
the first term $\partial _{\overline{A}}\bra{\psi(\overline{A})}\hat{T}%
\ket{\psi(A)}$ and $\bra{\psi(\overline{A})}\hat{T}\ket{\psi(A)}$ in
gradient Eq.~\eqref{eqn: gradient expression} can be represented as a
triple-layer tensor network. The second term in Eq.~\eqref{eqn: gradient
expression} can be similarly represented as a double-layer tensor network
without the layer of transfer operator $\hat{T}$.

By approximately contracting these 2D tensor network, one can evaluate the
cost function and its gradient and subsequently perform variational
optimization. In the following, we employ the split-CTMRG method \cite{Xu2025,naumann_2025} to
contract the multilayer network in Eqs.~\eqref{eqn: residual entropy iPEPS
expression} and~\eqref{eqn: gradient expression}.

\begin{table*}[t]
\caption{Configuration number per site $w$ obtained from variational iPEPS
calculations for different bond dimensions $D$ and $\protect\chi$. The upper
(lower) entry in each cell corresponds to the result for $w_c$ ($w_h$),
respectively.}%
\begin{ruledtabular}
\begin{tabular}{cccccccc}\label{Table: Residual entropy result from our method}
\diagbox[width=6em,innerleftsep=1.5em,innerrightsep=1.5em]{\hfill$\chi$}{D\hfill} & 2 & 3 & 4 & 5 & 6 & 7\\
\midrule
\multirow{2}{*}{30}
 & 1.5073979 & 1.5074407 & 1.5074432 & 1.5074439 & 1.5074439 & 1.5074439\\
 & 1.5074195 & 1.5074476 & 1.5074486 & 1.5074486 & 1.5074493 & 1.5074497\\
\midrule
\multirow{2}{*}{50}
 & 1.5073980 & 1.5074418 & 1.5074449 & 1.5074467 & 1.5074477 & 1.5074482\\
 & 1.5074195 & 1.5074525 & 1.5074529 & 1.5074531 & 1.5074533 & 1.5074538\\
\midrule
\multirow{2}{*}{70}
 & 1.5073980 & 1.5074429 & 1.5074461 & 1.5074482 & 1.5074503 & 1.5074511\\
 & 1.5074195 & 1.5074539 & 1.5074550 & 1.5074552 & 1.5074552 & 1.5074557\\
\midrule
\multirow{2}{*}{100}
 & 1.5073981 & 1.5074446 & 1.5074477 & 1.5074497 & 1.5074521 & 1.5074532\\
 & 1.5074195 & 1.5074553 & 1.5074572 & 1.5074576 & 1.5074577 & 1.5074578\\
\midrule
\multirow{2}{*}{150}
 & 1.5073981 & 1.5074454 & 1.5074486 & 1.5074510 & 1.5074528 & 1.5074533\\
 & 1.5074195 & 1.5074554 & 1.5074582 & 1.5074584 & 1.5074584 & 1.5074584\\
\end{tabular}
\end{ruledtabular}
\end{table*}

Let us briefly introduce this method. As shown in Figs.~\ref{Fig:Variational
scheme}(b) and~\ref{Fig:Variational scheme}(d), the central idea is to
compress the triple-layer network into a multi-site, single-layer network
comprised by the local tensors $\{A,O,\overline{A}\}$, while avoiding the
direct contraction over the physical indices between each layer. The
resulting infinite single-layer network is approximated by a set of corner
transfer matrix (CTM) environment tensors, including the corner tensors $%
\{C_{i}\}$ and the edge tensors $\{T^{\text{ket}}_{i}\},\{T^{\text{int}%
}_{i}\}, \{T^{\text{bra}}_{i}\}$ where $(i=1-4)$, corresponding to the
ket/int/bra layers of the triple-layer structure, respectively. These
environment tensors are obtained by iteratively absorbing the local tensors $%
\{A,O,\overline{A}\}$ and applying truncations via projectors. Further
details of the algorithm can be found in Ref.~\cite{Xu2025}. Once the CTM
tensors are converged, two contraction terms shown in Figs.~\ref%
{Fig:Variational scheme}(a) and ~\ref{Fig:Variational scheme}(c), involved in calculating residual
entropy and the corresponding gradient, can be calculated as depicted in
Figs.~\ref{Fig:Variational scheme}(b) and~\ref{Fig:Variational scheme}(d). Compared to the traditional
variational tensor network approaches, the split-CTMRG method exploits the
internal structure of the triple-layer network, thereby reducing
computational cost and enabling the optimization of iPEPS with larger bond
dimension $D$ and correspondingly more accurate estimates of the residual
entropy.

For the calculation of $S_h$ and $S_c$, the local tensor $A$ of the
fixed-point iPEPS is set to be real for both the transfer operators of $I_h$
and $I_c$, ensuring that the cost function remains real during optimization.
In addition, no symmetry constraints are imposed on the local tensor $A$, and
the directional move split-CTMRG scheme~\cite{Xu2025} is employed in the
optimization process. The optimization of the iPEPS tensor $A$ is carried
out using the BFGS algorithm, with convergence criteria set to $\|g\|_\infty
\approx 10^{-5}$-$10^{-6}$. The numerical error of our method is
estimated to be at the order of $5\times 10^{-7}$, with further details
available in the dataset \cite{data}. The numerical results are summarized in
Table~\ref{Table: Residual entropy result from our method}, where the
configuration number per site is reported for iPEPS bond dimensions $%
(D=2-7)$ and CTM bond dimensions $(\chi=30-150)$.

Our results are in agreement with the most recent Monte-Carlo results
presented in Table~\ref{Table: MC results}, as well as with previous tensor
network results \cite{Vanderstraeten2018}. Notably, our results for
hexagonal ice $I_h$ with bond dimension $D=2-4, \chi=100$ are slightly
higher than the values reported in \cite{Vanderstraeten2018} with same bond
dimension $D$: [$w=1.50735(D=2),1.507451(D=3), 1.507456(D=4)$]. This
discrepancy arises as no symmetry constraint is imposed on $A$ in our
method, which allows for a larger variational space to be explored. As the
bond dimension increases, the difference between $w_c$ and $w_h$ shows a
tendency of decrease. For the largest bond dimensions considered $%
(D=7,\chi=150)$, the relative difference between $w_h$ and $w_c$ is at the
order of $5\times 10^{-6}$.

To mitigate finite-$D$ and finite-$\chi$ effects, we perform an
extrapolation to the limit $D,\chi \to \infty$, based on the scaling
hypothesis proposed in Ref.~\cite{vanhecke_2022}. Within this framework, the
dependence on finite $D$ and $\chi$ is captured by an effective correlation
length, such that
\begin{equation}
S(D,\chi)=S(\xi(D,\chi)).
\end{equation}
For each pair of $(D,\chi)$, the correlation length can be extracted from
the optimized iPEPS tensor $A$ and the corresponding split-CTMRG environment
tensors. The corresponding details are included in Appendix~\ref%
{Appendix:Extracting correlation length from split-CTMRG method}.

\begin{figure}[t]
\centering
\includegraphics[width=1\linewidth]{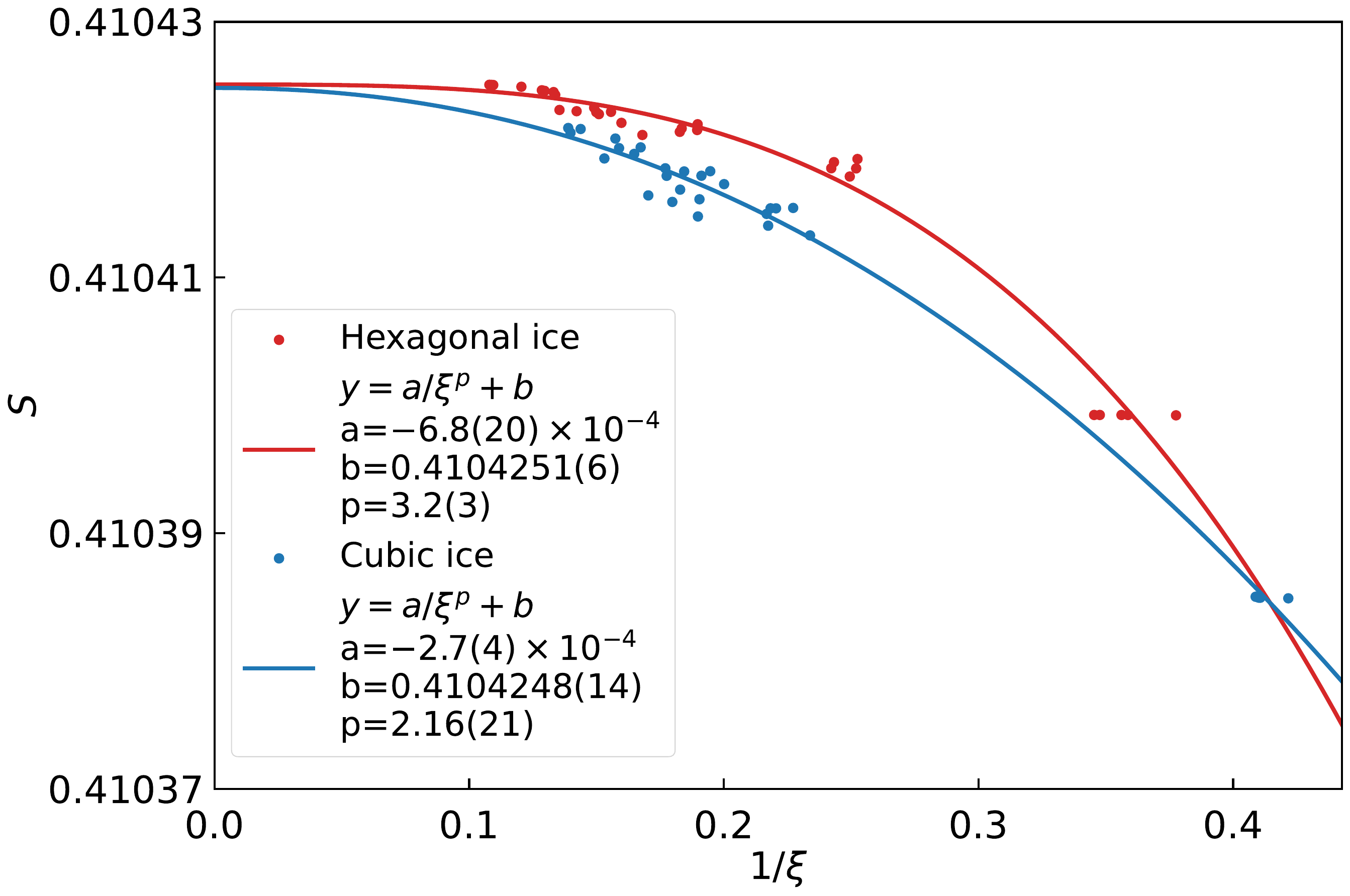}
\caption{Extrapolation for the results of residual entropy with the fitting
function $S(\protect\xi)=a/\protect\xi^p+S_\infty$.}
\label{Fig:residual entropy result}
\end{figure}

Figure~\ref{Fig:residual entropy result} presents the data collapse under this
hypothesis for residual entropy of hexagonal ice and cubic ice. We perform
an extrapolation for $\xi\to\infty$ with the fitting equation $%
S(\xi)=a/\xi^p+S_\infty$. The fitting result is $S_h=0.4104251(6)$ and $%
S_c=0.4104248(14)$, indicating that $S_c$ and $S_h$ are equal within the
numerical accuracy of our method.

\section{Conclusion and discussion}

\label{Sec:: Conclusion and discussion} In this work, we employed tensor
network methods to study and compare the residual entropies of hexagonal ice
$I_h$ and cubic ice $I_c$. Our results provide strong numerical evidence
that the two residual entropies are equal within the accuracy of our
approach.

First, we numerically confirm the normality of the transfer operator $\hat{M}
$ of cubic ice by calculating the fidelity $\mathcal{F}$ between $\hat{M}%
\hat{M}^T$ and $\hat{M}^T\hat{M}$. Within the tensor network framework, this
fidelity is represented as a four-layer infinite 2D tensor network. By
contracting this network using the VUMPS algorithm, we obtained $\mathcal{F}%
\approx0.999903$, indicating that $\hat{M}$ is highly close to a normal
operator. This high degree of normality provides strong support for the
applicability of variational tensor network methods without additional
symmetry constraints, consistent with the observation that $S_h$ and $S_c$
coincide.

Second, we directly computed the residual entropy in the thermodynamic limit
using a variational iPEPS optimization method. Without imposing any symmetry
constraints on the local tensor $A$, we find that the difference between $S_h
$ and $S_c$ is indistinguishable within numerical accuracy.

Generally speaking, the analysis of normality can be formulated as checking
whether two tensor network states constructed from different local tensors
are identical. While fidelity provides a global diagnostic, it is also
natural to address this problem at the level of local tensor relations. In
one dimension, such equivalence is well characterized by the fundamental
theorem of matrix product states or operators (MPS/MPO). However, for
two-dimensional PEPOs, no comparably general criterion is available and this
problem remains to be further explored. A more detailed discussion is
presented in Appendix.~\ref{Appendix:Normality examination from local tensor
relations}.

Looking ahead, recent theoretical and experimental studies have revealed a
rich family of ice phases, including more than 20 phases with distinct
structures \cite%
{Salzmann_2019,Bartels-Rausch_2012,gasser_2021,yamane_2021,salzmann_2021,Zhang_2025,rescigno_2025}%
. In these phases the oxygen atoms form regular lattice different from those
in hexagonal ice and cubic ice, while maintaining a coordination number of
4, with hydrogen atoms located along the bonds between oxygen atoms.
Consequently, the residual entropy arising from the extensive number of
configurations that satisfy the ice rules can be systematically investigated
in these new ice phases, which remain largely unexplored. The method
presented in this work provides a promising approach for computing the
residual entropy of these ice structures.

Furthermore, in the study of three-dimensional statistical physics, the
complex stacking of layers in three-dimensional structures introduces
additional challenges, where the layer-to-layer transfer matrix of the
system is not guaranteed to be Hermitian. In such non-Hermitian cases, the
perspective of normality and the numerical methods presented in this paper
will be valuable for investigating these issues.

\textbf{Acknowledgments.} This research is supported by the National Key
Research and Development Program of China (Grant No. 2023YFA1406400).

\textbf{Data availability.} The data that support the findings of this
article are openly available \cite{data}.

\appendix

\section{Extracting correlation length from split-CTMRG method}

\label{Appendix:Extracting correlation length from split-CTMRG method} 
\begin{figure}[t]
\includegraphics[width=1\linewidth]{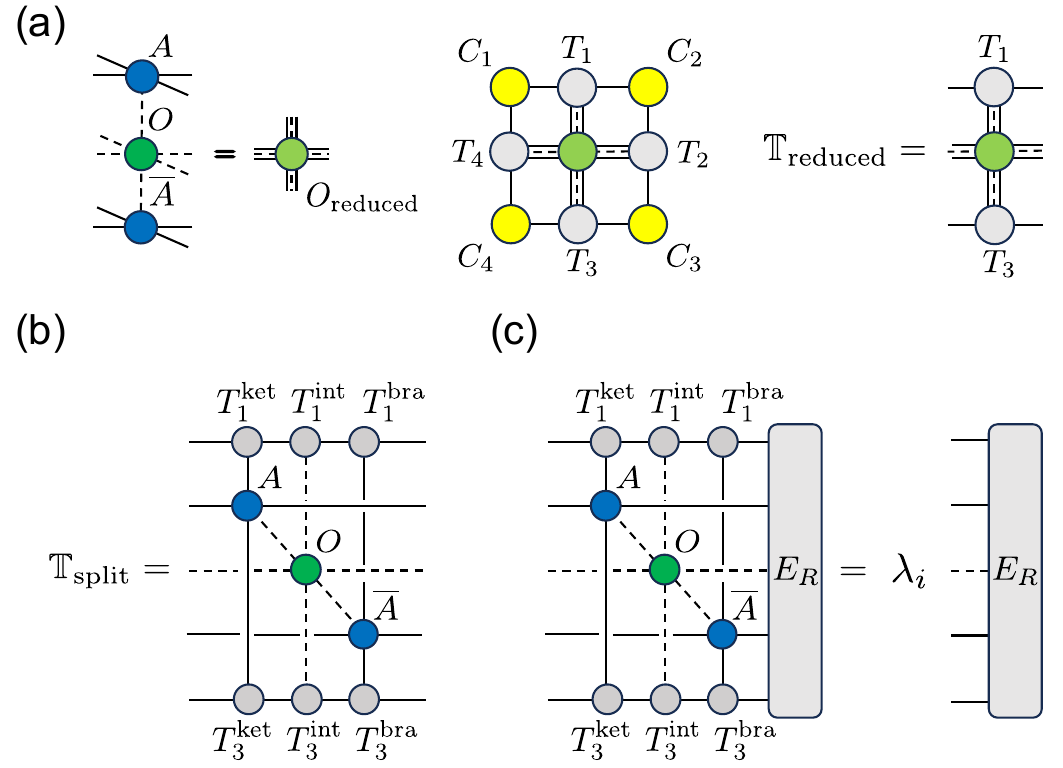}
\caption{Channel operators and extraction of correlation length. (a) The CTM
environment tensors for the traditional reduced method and the channel
operator $\mathbb{T}_\text{reduced}$. (b) Channel operator $\mathbb{T}_\text{%
split}$ for the split-CTMRG scheme. (c) Eigenvalue equation for the channel
operator $\mathbb{T}_\text{split}$. }
\label{Fig:Correlation length}\centering
\end{figure}

With
the optimized iPEPS fixed point $\ket{\psi(A)}$ of the transfer operator $%
\hat{T}$, the correlation length can be extracted from the channel operator $%
\mathbb{T}$ composed of local tensors $\{A,O,\overline{A}\}$ and the
effective environment tensors. In the traditional CTMRG framework, the
triple-layer network is reduced to a single-layer structure and the
effective environment is approximated by the corner tensors $\{C_i\}$ and
edge tensors $\{T_i\}$. Correspondingly, the channel operator $\mathbb{T}_%
\text{reduced}$ is defined as shown in Fig.~\ref{Fig:Correlation length}(a).
This definition can be generalized to the split-CTMRG method employed in
this work as shown in Fig.~\ref{Fig:Correlation length}(b), where the
environment edge tensors are split into different layers $\{T^{\text{ket}%
}_{i}\},\{T^{\text{int}}_{i}\}, \{T^{\text{bra}}_{i}\}$. By solving the
eigenvalue equation of the channel operator $\mathbb{T}_\text{split}\ket{E_R}%
=\lambda_i\ket{E_R}$, as illustrated in Fig.~\ref{Fig:Correlation length}%
(c), the correlation length can be extracted as
\begin{equation}
\xi=-\frac{1}{\ln|\lambda_1/\lambda_0|},
\end{equation}
where $\lambda_0$ and $\lambda_1$ are the leading and subleading
eigenvalues.

\section{Normality examination from local tensor relations}

\label{Appendix:Normality examination from local tensor relations}

\begin{figure}[t]
\centering
\includegraphics[width=1\linewidth]{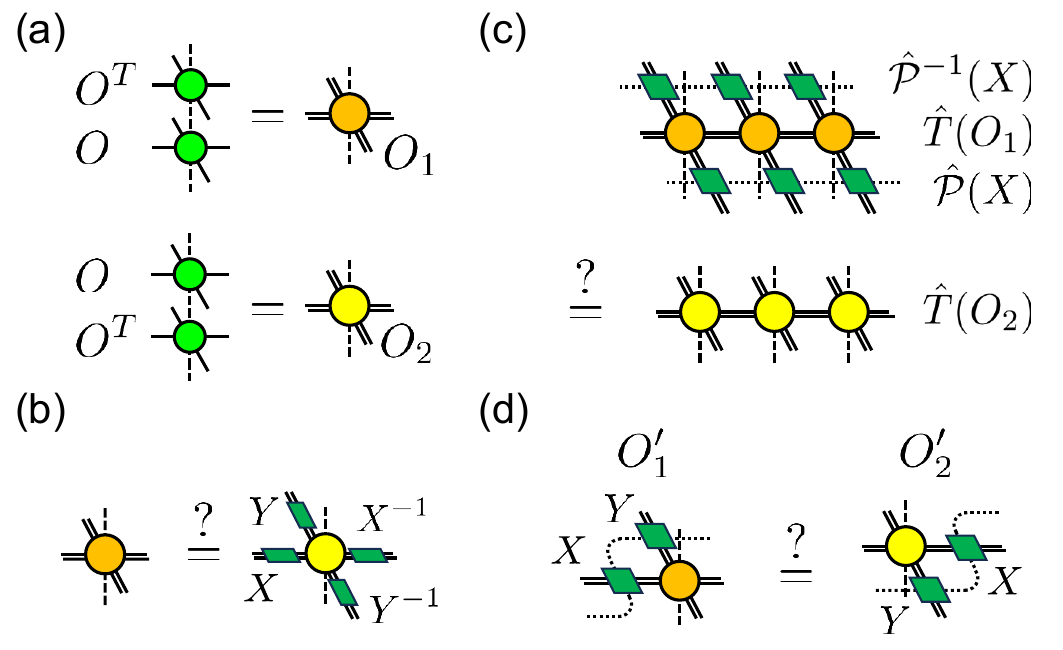}
\caption{Possible gauge transformation between local tensors of $\hat{M}\hat{%
M}^T$ and $\hat{M}^T\hat{M}$. (a) Local tensor $O_1$ ($O_2$) forming the
transfer operators $\hat{M}\hat{M}^T$ ($\hat{M}^T\hat{M}$). $O^T$ denotes
the tensor obtained by swapping the vertical indices of tensor $O$. (b)
Invertible matrix transformation induced by invertible matrices $X$ and $Y$
acting on the links. (c) Invertible MPO transformation induced by invertible
MPO $\hat{\mathcal{P}}(X)$. (d) Pulling-through transformation induced by
local tensor $X$ and $Y$.}
\label{Fig:PEPO_relation_check}
\end{figure}

As the numerical results suggest that the transfer operator $\hat{M}$ is
highly normal, we further examine whether its normality can be established at the level of local tensors. Although the local
tensor $O_1$ of $\hat{M}\hat{M}^{T}$ and the local tensor $O_2$ of $\hat{M}%
^{T}\hat{M}$ [shown in Fig.~\ref{Fig:PEPO_relation_check}(a)] are not
identical, the PEPO structure admits certain gauge transformations under
which distinct local tensors may represent the same PEPO. To this end, we
numerically analyze three typical gauge transformations that provide
sufficient conditions for the desired equality $\hat{M}\hat{M^T}=\hat{M^T}%
\hat{M}$.

\subsection{Invertible matrix transformation} We first examine whether the
two local tensors can be related by an invertible local matrix
transformation, as illustrated in Fig.~\ref{Fig:PEPO_relation_check}(b). In
this scenario,
\begin{equation}
O_1 = \left( X^{-1} \otimes Y^{-1} \right) O_2 \left( X \otimes Y \right),
\end{equation}
where $X $ and $Y $ denote invertible matrices acting on the links.

\subsection{Invertible MPO transformation} We next investigate whether the
row-to-row transfer matrices constructed from the two tensors can be related
by an invertible matrix product operator (MPO), as illustrated in Fig.~\ref%
{Fig:PEPO_relation_check}(c). In this setup,
\begin{equation}
\hat{T}(O_1) = \hat{\mathcal{P}}(X) \, \hat{T}(O_2) \, \hat{\mathcal{P}}%
(X^{-1}),
\end{equation}
where $\hat{T}(O_1) $ and $\hat{T}(O_2) $ denote the respective row-to-row
transfer matrices, and $\hat{\mathcal{P}}(X) $ is an invertible MPO whose
local tensor is $X $.

\subsection{Pulling-through transformation} Moreover, the MPO relating the
two PEPOs need not be aligned with the row or column directions but may
instead pass through the PEPO along more general trajectories, as
illustrated in Fig.~\ref{Fig:PEPO_relation_check}(d). This condition is
known as the ``pulling-through condition,'' which has been extensively
discussed in the context of topological order \cite%
{Williamson_2016,bultinck_2018,cirac_2021,Molnar_2018,sahinoglu_2021}.

We examine these possibilities by minimizing the fidelity between the two
resulting tensors after gauge transformation. For all three cases, our
numerical results show that the fidelity does not converge to one,
indicating that $O_1$ and $O_2$ cannot be made identical at the level of
both local tensors and row-to-row transfer matrices through such local
matrix/MPO gauge transformations. These results suggest that the high degree
of normality of $\hat{M}$ can only be established at the level of
two-dimensional thermodynamic limits and cannot be related by local matrix
transformations nor one-dimensional MPO transformations.

More generally, this problem can be formulated as checking whether two
tensor network states are identical. In the case of matrix product states or
operators (MPS/MPO), the fundamental theorem of MPS provides a powerful
criterion \cite{CIRAC_2017}: Two translational invariant MPS $|\psi
(A)\rangle $ and $|\psi (B)\rangle $ are equal if and only if there exists
an invertible matrix $X$ such that $A=XBX^{-1}$. Therefore, to prove (or
disprove) the equality of the two MPSs, it suffices to determine whether such
an invertible matrix $X$ exists. However, the situation of PEPO/PEPS is
considerably more complicated. It has been proved that no fundamental
theorem exists for general PEPS \cite{scarpa_2020}. Instead, fundamental
theorem of PEPS has only been established for certain special subclasses of
PEPS that satisfy additional injectivity conditions \cite{cirac_2021}, such
as semi-injectivity \cite{Molnar_2018}. These injectivity conditions were
primarily developed in the study of quantum states with topological order.
For PEPOs arising from statistical mechanics, to the best of our knowledge,
no fundamental theorem has been established that specifically addresses this
subclass of PEPOs. Consequently, the known versions of the PEPS fundamental
theorem do not apply directly to our case and we are unable to find any
necessary and sufficient condition that would directly prove or disprove the
desired equality. The analysis of this high degree of normality from local
tensor perspective and, more generally, the fundamental theorem for PEPOs
arising from statistical mechanics remains a subject for future exploration.

\end{document}